\RequirePackage{plautopatch} 
\documentclass{article}
\usepackage[utf8]{inputenc}

\usepackage{url}
\usepackage{listings,jvlisting}
\usepackage{amssymb}
\usepackage{tabularx}
\usepackage{amsmath}
\usepackage{mdframed}
\usepackage{graphicx}
\usepackage{setspace}
\usepackage{here}
\usepackage{authblk}
\usepackage{mathtools}
\usepackage{amsthm}
\usepackage{algorithmic}
\usepackage{algorithm}
\usepackage{cite}

\def\be{{\beta}}

\def\de{{\delta}}

\def\la{{\lambda}}
\def\La{{\Lambda}}

\def\th{{\theta}}

\def\bbe{{\text{\boldmath $\beta$}}}

\def\bth{{\text{\boldmath $\theta$}}}

\def\btau{{\text{\boldmath $\tau$}}}

\def\bbeh{{\hat \bbe}}
\def\beh{{\hat \be}}

\def\thh{{\hat \th}}

\def\tauh{{\hat \tau}}

\def\Lah{{\hat \La}}

\def\bbeh{{\widehat \bbe}}

\def\bthh{{\widehat \bth}}

\def\btauh{{\widehat \btau}}

\def\Ga{{\Gamma}}

\def\La{{\Lambda}}

\def\x{{\text{\boldmath $x$}}}

\def\vh{{\hat v}}

\def\thh{{\hat \th}}

\begin{document}
\title{Frailty Model with Change Point for Survival Analysis}
\author[1,2]{Masahiro Kojima\footnote{Address: Biometrics Department, R\&D Division, Kyowa Kirin Co., Ltd.
Otemachi Financial City Grand Cube, 1-9-2 Otemachi, Chiyoda-ku, Tokyo, 100-004, Japan. Tel: +81-3-5205-7200 \quad
E-Mail: masahiro.kojima.tk@kyowakirin.com}}
\author[3]{Shunichiro Orihara}
\affil[1]{Kyowa Kirin Co., Ltd}
\affil[2]{The Institute of Statistical Mathematics}
\affil[3]{Tokyo Medical University}
\maketitle
\abstract{\noindent
We propose a novel frailty model with change points applying random effects to a Cox proportional hazard model to adjust the heterogeneity between clusters. Because the frailty model includes random effects, the parameters are estimated using the expectation-maximization (EM) algorithm. Additionally, our model needs to estimate change points; we thus propose a new algorithm extending the conventional estimation algorithm to the frailty model with change points to solve the problem. We show a practical example to demonstrate how to estimate the change point and random effect. Our proposed model can be easily analyzed using the existing R package. We conducted simulation studies with three scenarios to confirm the performance of our proposed model. We re-analyzed data of two clinical trials to show the difference in analysis results with and without random effect. In conclusion, we confirmed that the frailty model with change points has a higher accuracy than the model without the random effect. Our proposed model is useful when heterogeneity needs to be taken into account. Additionally, the absence of heterogeneity did not affect the estimation of the regression coefficient parameters.
}
\par\vspace{4mm}
\section{Introduction}
\label{sec1}
Cox proportional hazard model is one of the primary analysis methods for time-to-event data. However, in recent years, there has been a case in which the assumption of proportional hazard does not hold in time-to-event data in some trials of immune checkpoint inhibitors such as nivolumab\cite{shitara2022nivolumab}. Immune checkpoint inhibitor (test drug group) shows a survival curve similar to the control group because events occur in a population less likely to respond to the test drug up to a point. On the other hand, after the point, the decrease in the survival probability of the test drug group became slower; the survival curve of the test drug group then moves away from that of the control group. Therefore, the assumption of proportional hazard may hold about each survival curve before and after the point. We call the point “{\it change point}” in this paper. 

To analyze these events accurately, the Cox proportional hazard model with change points is a valuable method.
Various types of research have been conducted on the Cox proportional hazard model with change points\cite{liang1990cox,pons2002estimation,liu2008monte,he2013sequential,xu2014bootstrapping,wang2021change,ozaki2022information}. Liang et al. (1990)\cite{liang1990cox}  proposed the proportional hazard model with one change point. Pons (2002)\cite{pons2002estimation} extended Liang's model to time-dependent covariates and proved the consistency of the estimators of coefficient parameters and the estimated change point. Liu et al. (2008)\cite{liu2008monte} and He et al. (2013)\cite{he2013sequential} proposed the maximal score tests for detecting change points using a simple Monte Carlo approach. Ozaki and Ninomiya (2022)\cite{ozaki2022information} proposed a novel information criterion to determine the number of the best change points.

Cluster effects are commonly assumed in various research fields (McNeish and Kelley, 2019 \cite{mcneish2019fixed}). For instance, in clinical research, there are clusters of primary diseases, clinical facilities, and severity of interest disease. Commonly, the cluster effects are included in assumed statistical models as random effects. The random effects capture cluster specific variation, and fixed effects (e.g. interested treatment effects) which are not affected by the clusters can be estimated more precisely. A frailty model is a Cox proportional hazard model that accounts for heterogeneity between clusters. The frailty model has various previous works (c.f. Klein (1992) \cite{klein1992semiparametric}; Vaida and Xu, 2000 \cite{vaida2000proportional}), however, change points has not yet to be considered.

In this study, we propose a novel frailty model with change points applying random effects to a Cox proportional hazard model to adjust the heterogeneity between clusters. Because the frailty model includes random effects, the parameters are estimated using the expectation-maximization (EM) algorithm proposed by Klein (1992) \cite{klein1992semiparametric}. Additionally, our model needs to estimate change points. We propose a new algorithm extending the conventional estimation algorithm to the frailty model with change points to solve the problem. Our proposed model can be easily analyzed using the existing R package. We show a practical example that shows how to estimate the change point and random effect. To confirm the performance of our proposed model, we conduct simulation studies with three scenarios. In addition, we re-analyzed data from two clinical trials to show the difference in the results with and without random effect. 

This paper is organized as follows. Section 2 describes the proposed frailty model with change point and introduces the estimation method of the model. Section 3 presents a practical example to demonstrate how to estimate the change point and random effect of proposed model. Section 4 describes the setting and results of the computer simulations. Section 5 presents the results of analysis of data from two published clinical trials. Section 6 provides a discussion of the results. The R program files used to analyze the simulation results and clinical trials are included in the Supplemental Material.

\section{Methods}\label{sec2}
We assume that there are $M(\geq 2)$ clusters, the sample size is $N$, $Y_{mi}$ is a time-to-event of subject $i$ for $m$-th cluster, and $\x_{mi}$ is a vector of $q$-dimensional covariates for $m$-th cluster. $Y_{mi}$ can be right censored; the observation data $T_{mi}$ is min$(Y_{mi},C_{mi})$, where $C_{mi}$ is a censoring time. We assume that $C_{mi}$ is independent of the other random variables that is the special case of the type $\rm\,I\,$ censoring (see Kalbfleisch and Prentice, 2002 \cite{kalbfleisch2011statistical}). We assume that there are $K$ change points.
The frailty model with change points considered in this paper is defined by giving coefficient parameters between each change point,
\begin{align}
\la(t_{mi};\x_{mi},\bbe_k,v_{km},\tau_k)=\la_0(t_{mi})v_{km}\exp\left[\bbe_k^T\x_{mi}\right]I(\tau_{k-1}<t_{mi}\leq\tau_k),
\end{align}
where 
$\la_0(\cdot)$ is a nonparametric hazard function, $v_{km}$ is a random effect on $m$-th cluster for interval $(\tau_{k-1},\tau_k]$, $\bbe_k$ is a vector of $q$-dimensional parameters for interval $(\tau_{k-1},\tau_k]$, and $\tau_k$ is an unknown change point for time point $k$. The $K$ change points hold $0=\tau_0<\tau_1<\cdots<\tau_K<\tau_{K+1}=T$, in which $T$ is the follow-up period. We assume that the distribution of $v_{km}$ is the gamma distribution with the shape $\frac{1}{\th_k}$ and the rate $\frac{1}{\th_k}$. Under this setting, the mean and variance of $v_{km}$ are $1$ and $\th_k$, respectively. To simplify the notation, let $\bbe=(\bbe_1^T,\ldots,\bbe_K^T,\bbe_{K+1}^T)^T$, $\btau=(\tau_1,\ldots,\tau_K)^T$, and $\bth=(\th_1,\ldots,\th_K,\th_{K+1})^T$.

From here, we consider a proposed estimating procedure. 
Under the settings, the likelihood function becomes
\begin{align}
    \label{eq:lf}
l(\la_0,\bbe,\btau,\th)&=l_1(\la_0,\bbe,\btau)+\sum^K_{k=1}l_2(\th_k).
\end{align}
where 
\begin{align}
l_1(\la_0,\bbe,\btau)&=\sum^{K}_{k=1}\sum^M_{m=1}\sum^{N_m}_{i=1}\Biggl[\de_{mi}\left\{\log(\la_0(t_{mi}))+\bbe_k^T\x_{mi}\right\}-v_{km}\exp(\bbe_k^T\x_{mi})\left(\La_0(t_{mi})-\La_0(\tau_{k-1})\right)\Biggr]I(\tau_{k-1}<t_{mi}\leq\tau_k),
\end{align}
and
\begin{align}
    l_2(\th_k)=-M\left(\frac{1}{\th_k}\log(\th_k)+\log\left(\Ga\left(\frac{1}{\th_k}\right)\right)\right)+\sum^M_{m=1}\left\{\left(\frac{1}{\th_k}+D_{km}-1\right)\log(v_{mk})-\frac{v_{mk}}{\th_k}\right\}.
\end{align}
where $N_m$ is the sample size on the $m$-th cluster, $\de_{mi}=I(T_{mi}=Y_{mi})$ and $D_{km}=\sum_{i\in E_k}\de_{mi}$, $E_k$ is the set of subject number to which the time-to-event in interval $(\tau_{j-1},\tau_j]$. $l_1(\la_0,\bbe,\btau)$ is the likelihood function for the Cox proportional hazard model and change points. $l_2(\th_k)$ is the likelihood function for the frailty.
Since the parameters cannot be estimated analytically because of the random effects, we consider the extension of Klein (1992)[9]. Moreover, because Klein's algorithm is intuitively straightforward, the extension is simple to apply to our proposed frailty model with change points.
The EM algorithm is as follows. First, to deal with the unobserved random effect, the E-step calculates the expected values of the random effect.

\bigskip
\noindent
{\bf The E-step}\\
The conditional distribution on the observed data is the gamma with shape parameter $A_{km}=\frac{1}{\th_k}+D_{km}$ and rate parameter $B_{km}=\frac{1}{\th_k}+\sum_{i\in E_k}\La_0(t_{mi})\exp(\bbe_k\x_{mi})I(\tau_{k-1}<t_{mi}\leq\tau_k)$. The conditional likelihood function is obtained by replacing $v_{km}$ in the likelihood function with $\frac{A_{km}}{B_{km}}$. The initial value of $\th_k$ is 1. The initial value of $\bbe_k$ is the estimator of the standard Cox proportional hazard model with change points.

\bigskip
\noindent
{\bf The M-step}\\
In the M-step, the estimator of the parameters that maximizes the conditional likelihood function in the E-step is computed. For the conditional likelihood function of $\bbe$ and $\btau$, the following profile likelihood function can be given through the estimation of nonparametric hazards $\Lah$.
\begin{align}
    \Lah_0(t_{mi})=\sum_{r:t_{mr}\leq t_{mi}}\frac{d_{mr}}{\sum_{j:t_{mi}<t_{mj}}\vh_{km}\exp(\bbe_k\x_{mj})I(\tau_{k-1}<t_{mj}\leq\tau_k)},
\end{align}
$d_{mr}$ is the number of event at $t_{mr}$.
$\bbe$ and $\btau$ are estimated using the following partial likelihood function,
\begin{align}
    \label{eq:be}
    pl_1(\bbe,\btau)=\sum^{k}_{k=1}\sum^M_{m=1}\sum^{N_m}_{i=1}\Biggl[\de_{mi}\bbe_k^T\x_{mi}-d_{mi}\log\left(\sum_{j:t_i<t_j} \vh_{km}\exp(\bbe_k^T\x_{jm})\right)\Biggr]I(\tau_{k-1}<t_j\leq\tau_k),
\end{align}
where $\vh_{km}=\frac{A_{km}}{B_{km}}$.
The estimator of $\th_k$ is obtained by maximizing the conditional likelihood function on the observed data, 
\begin{align}
    \label{eq:th}
    cl_2(\th_k)=-M\left(\frac{1}{\th_k}\log(\th_k)+\log\left(\Ga\left(\frac{1}{\th_k}\right)\right)\right)+\sum^M_{m=1}\left\{\left(\frac{1}{\th_k}+D_{km}-1\right)(\log(A_{km})-\log(B_{km}))-\frac{A_{km}}{\th_k B_{km}}\right\}
\end{align}
The candidate of change point $\tau_k$ is chosen from each time-to-event $Y_{mi}$ because the change point is considered as the event point that may change the slope of the Cox regression in this model. Hence, for each $\tau_k$, the actual time-to-event is input. $\bbe$ and $\bth$ are then calculated. We fit all time-to-event combinations to the change point $\tau_k$ and compute estimators $\bbeh$ of $\bbe$ from the equation (\ref{eq:be}) and $\bthh$ of $\bth$ from the equation (\ref{eq:th}) for each fixed change point. The maximum likelihood estimator is the combination of $\bbeh$, $\btauh$, and $\bthh$ that maximizes the likelihood function (\ref{eq:lf}), where $\btauh$ is input actual times-to-events.

\section{Practical example}
We demonstrate the analysis of change points. We create the example dataset shown in Table \ref{ex_dat}.
\begin{table}[H]
  \begin{center}
\caption{Example dataset\label{ex_dat}}
\begin{tabular}{|c|c|c|c|c|c|c|c|}\hline
\multicolumn{4}{|c|}{Placebo Group}&\multicolumn{4}{c|}{Treatment Group}\\\hline
ID& ST & Censor & Cluster & ID& ST & Censor & Cluster\\ \hline
1	&	10	&	Yes & 1 &	16	&	10	&	Yes & 1	\\
2	&	25	&	No	& 1&	17	&	15	&	Yes & 1	\\
3	&	30	&	No	& 1&	18	&	25	&	No & 1	\\
4	&	45	&	No	& 2&	19	&	40	&	No & 2	\\
5	&	50	&	No	& 2&	20	&	45	&	No & 2	\\
6	&	55	&	Yes	& 2&	21	&	60	&	Yes & 2	\\
7	&	60	&	No	& 3&	22	&	65	&	Yes & 3	\\
8	&	65	&	No	& 3&	23	&	70	&	No & 3	\\
9	&	70	&	Yes	& 3&	24	&	75	&	No & 3	\\
10	&	75	&	No	& 1&	25	&	80	&	Yes & 1	\\
11	&	80	&	No	& 1&	26	&	85	&	No & 1	\\
12	&	85	&	No	& 2&	27	&	90	&	Yes	 & 2\\
13	&	90	&	Yes	& 2&	28	&	95	&	No	 & 2\\
14	&	95	&	No	& 3&	29	&	100	&	Yes & 3	\\
15	&	100	&	Yes	& 3&	30	&	100	&	Yes	 & 3\\\hline
\end{tabular}
\\
      \footnotesize{ST: survival time (Week)}
  \end{center}
\end{table}
First, we introduce the analysis of a single change point and no random effect. The set of candidate change points consists of actual times-to-event. Data for the 12 candidate points of change point $\tau_1$ are $(25,30,40,45,50,60,65,70,75,80,85,95,100)$. The partial likelihood function with the unknown parameter $\bbe=(\be_1,\be_2)^T$ for each $\tau_j$ is
\begin{align}
    pl(\bbe,\tau_1)=&\sum_{i=1}^{30}\left\{(I(0< t_i\leq\tau_1)\be_1+I(\tau_1< t_i\leq 100)\be_2)x_i\right\}\nonumber\\
    &\hspace{0.5cm}-\log\left(\sum_{m\in R_1}\exp(\be_1x_m)\right)-\log\left(\sum_{m\in R_2}\exp(\be_2x_m)\right).
\end{align}
$R_1$ is the risk set for $t_i\geq 0$. However, when $t_i\geq\tau_1$, the data are treated as censored data, $R_2$ is the risk set for $t_i\in [\tau_1,100]$. For each change point, we calculate the estimator $\bbeh$ maximizing the partial likelihood function. $\be_1$ and $\be_2$ can be computed independently. The combination of $\bbeh$ and $\tauh$ that maximizes the partial likelihood function is the maximum estimator. The maximum likelihood estimator was $\bbeh=(0.07, -0.76)$ and $\tauh=50$. Data analysis was performed using coxph function of the survival package in R. For detailed information on the algorithms, please refer to the supplemental material.

\begin{figure}[H]
  \begin{center}
  \includegraphics[width=15cm]{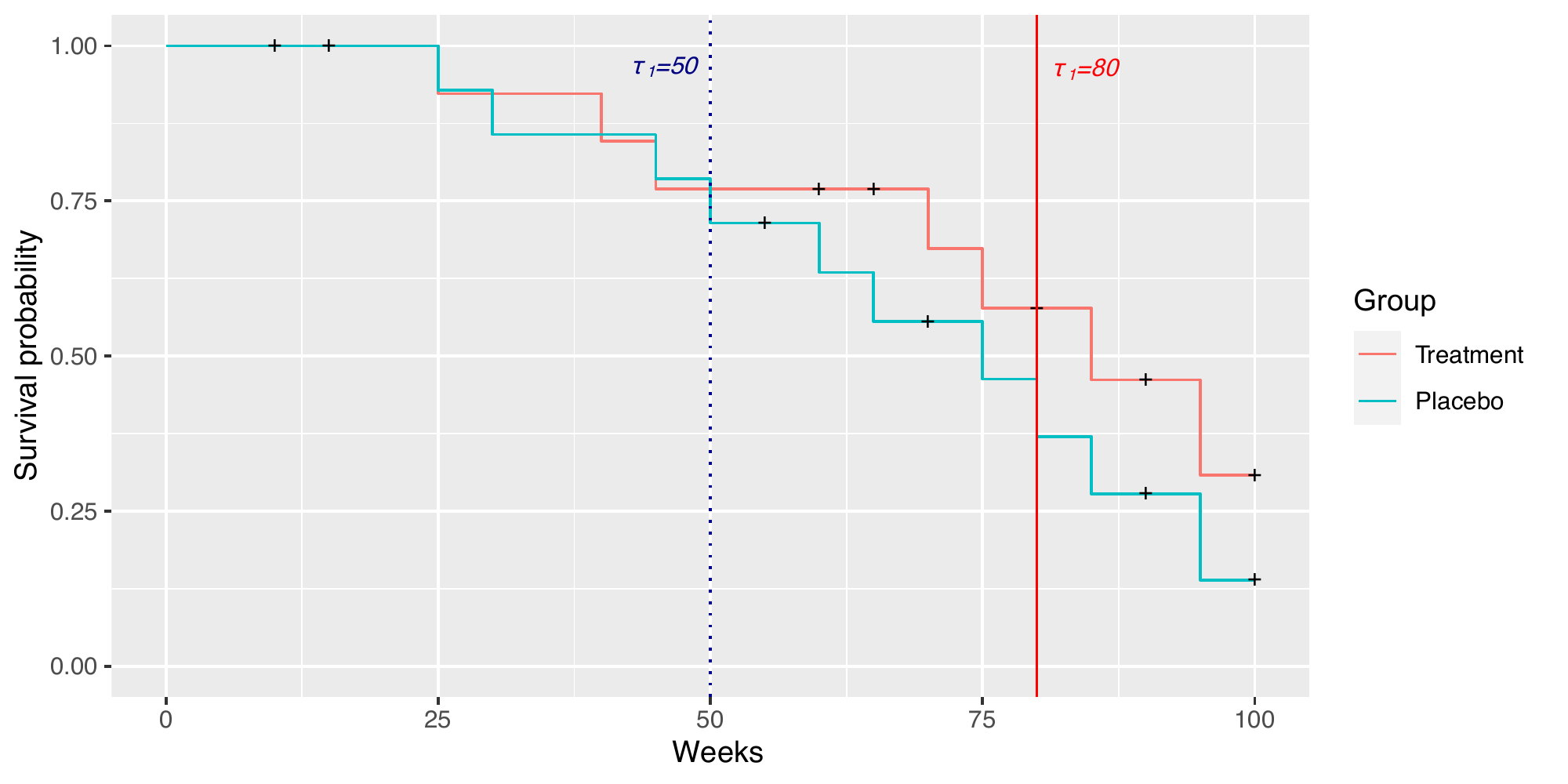}
  \caption{Survival curves and change points of example data}
  \label{ex_fig}
      \footnotesize{$+$ means censored. The red line and red letters are the change points of the frailty model with one change point. The dark blue dotted line and dark blue letters are the change points of the Cox proportional hazard with one change point.}
  \end{center}
\end{figure}

Next, we show the analysis results of the frailty model with one change point. The partial likelihood is 
\begin{align}
    pl_1(\bbe,\tau_1)=\sum^3_{m=1}\sum^{N_m}_{i=1}\Biggl[\de_{mi}\bbe_k^T\x_{mi}-d_{mi}\log\left(\sum_{j:t_i<t_j} \vh_{km}\exp(\bbe_k^T\x_{jm})\right)\Biggr]I(\tau_{k-1}<t_j\leq\tau_k),
\end{align}
and
\begin{align}
    cl_2(\th_1)&=-3\left(\frac{1}{\th_1}\log(\th_1)+\log\left(\Ga\left(\frac{1}{\th_1}\right)\right)\right)+\sum^3_{m=1}\left\{\left(\frac{1}{\th_1}+D_{1m}-1\right)(\log(A_{1m})-\log(B_{1m}))-\frac{A_{1m}}{\th_1 B_{1m}}\right\},\nonumber\\
    cl_2(\th_2)&=-3\left(\frac{1}{\th_2}\log(\th_2)+\log\left(\Ga\left(\frac{1}{\th_2}\right)\right)\right)+\sum^3_{m=1}\left\{\left(\frac{1}{\th_1}+D_{2m
}-1\right)(\log(A_{2m})-\log(B_{2m}))-\frac{A_{2m}}{\th_2 B_{2m}}\right\}.    
\end{align}
We input $50$ to $\tau_1$, $D_{11}=3$, $D_{12}=4$, $D_{13}=0$, $D_{21}=3$, $D_{22}=2$, and $D_{23}=5$.
From the coxph function in R, the maximum likelihood estimators were easily calculated $\bbeh=(-0.35, -1.56)$, $\tauh=80$, and $\bthh=(0.00, 1.78)$. Note that $\thh_1$ was not 0 but a very small value. When survival time was biased by clusters, adjusting for the random effect resulted in different change point estimates compared with the no random effect model.
The survival curve for each group and change points are shown in Figure \ref{ex_fig}.
\section{Simulation Study}
We evaluated the performance of the frailty model with change points. We assume that the sample size is $500$, and the four category cluster is randomly assigned from $(1,2,3,4)$ with a probability of $0.25$, $x_{mi}$ is randomly assigned $0$ or $1$ with a probability of $0.5$, the change point is $250$, the follow-up period is $600$, $\bbe=(\be_0,\be_1)^T=(0,0.5)^T$, and the non-informative censors occur a probability of $0.1$. The number of simulations is 10,000. We prepared three scenarios. {\bf Scenario 1} has no frailty. The survival data are generated from $\exp(\be_0x_{im})\times Exponential\left(\frac{1}{300}\right)$ up to the change point and $\exp(\be_1x_{im})\times Exponential\left(\frac{1}{300}\right)$ after the change point. {\bf Scenario 2} has the frailty $v_{1m}$ and $v_{2m}$ generated from $Gamma(0.1,0.1)$. The survival data are generated from $v_{1m}\exp(\be_1x_{im})\times Exponential\left(\frac{1}{300}\right)$ up to the change point and $v_{2m}\exp(\be_2x_{im})\times Exponential\left(\frac{1}{300}\right)$ after the change point. {\bf Scenario 3} is the case in which the parameter of Gamma distribution in Scenario 2 changed to $0.2$. The evaluation index for the simulation is bias and mean squared error (MSE). The bias is the average of estimator for each simulation $-$ the true parameter value. The MSE is the average of the squared error of the difference between the estimators and true parameter values for each simulation. The simulation program is included in the supplemental material.    
\subsection{Simulation Results}
The simulation results are shown in Table \ref{sim_res}. When there was no random effect in Scenario 1, the estimated results of the random effects are close to zero in the frailty model. The MSEs for $\bbeh$ and $\tauh_1$ were smaller in the frailty model. In Scenarios 2 and 3, the MSE of the change point was improved by adjusting for the variation effect. The change point bias was larger for the frailty model.
\begin{table}[H]
  \begin{center}
\caption{Simulation results\label{sim_res}}
\begin{tabular}{|c|c|c|c|c|}\hline
Parameter&\multicolumn{2}{c|}{CP without random effect}&\multicolumn{2}{c|}{Frailty model}\\\hline
\multicolumn{5}{|l|}{Scenario 1 ($\th_1=\th_2=0$)}\\\hline
& Bias & MSE & Bias& MSE \\ \hline
$\be_1$&-0.067&0.045&-0.042&0.036\\
$\be_2$&0.071 &0.057 &0.022 &0.052\\
$\tau_1$&14.7 &7958.8 &25.7 &7921.4\\
$\th_1$& - & - &-0.010 &0.001\\
$\th_2$& - & - &-0.012 &0.001\\\hline
\multicolumn{5}{|l|}{Scenario 2 ($\th_1=\th_2=0.1$)}\\\hline
& Bias & MSE & Bias& MSE \\ \hline
$\be_1$	&-0.067 &0.047 &-0.023 &0.032\\
$\be_2$	&0.049 &0.055 &0.007 &0.049\\
$\tau_1$ &18.5 &8278.1 &19.0 &\textcolor{red}{\bf 6840.9}\\
$\th_1$	& - & - &0.086 &0.009\\
$\th_2$	& - & - &-0.016 &0.020\\\hline
\multicolumn{5}{|l|}{Scenario 3 ($\th_1=\th_2=0.2$)}\\\hline
& Bias & MSE & Bias& MSE \\ \hline
$\be_1$	&-0.067 &0.047 &-0.020 &0.030\\
$\be_2$	&0.026 &0.054 &-0.008 &0.046\\
$\tau_1$ &22.6 &8722.9 &26.6 &\textcolor{red}{\bf 6346.9}\\
$\th_1$	& - & - &0.179 &0.036\\
$\th_2$	& - & - &-0.039 &0.057\\\hline
\end{tabular}
\\
      \footnotesize{CP: Cox proportional hazard model with change point, Frailty model: Frailty model with change point. Red letter: MSE result is superior when the frailty is taken into account.}
  \end{center}
\end{table}

\section{Clinical Trials}
We show how the frailty model behaves when applied to data from two clinical trials.
\subsection{Re-analysis of data of a clinical trial on primary biliary cholangitis}
\begin{figure}[H]
  \begin{center}
  \includegraphics[width=15cm]{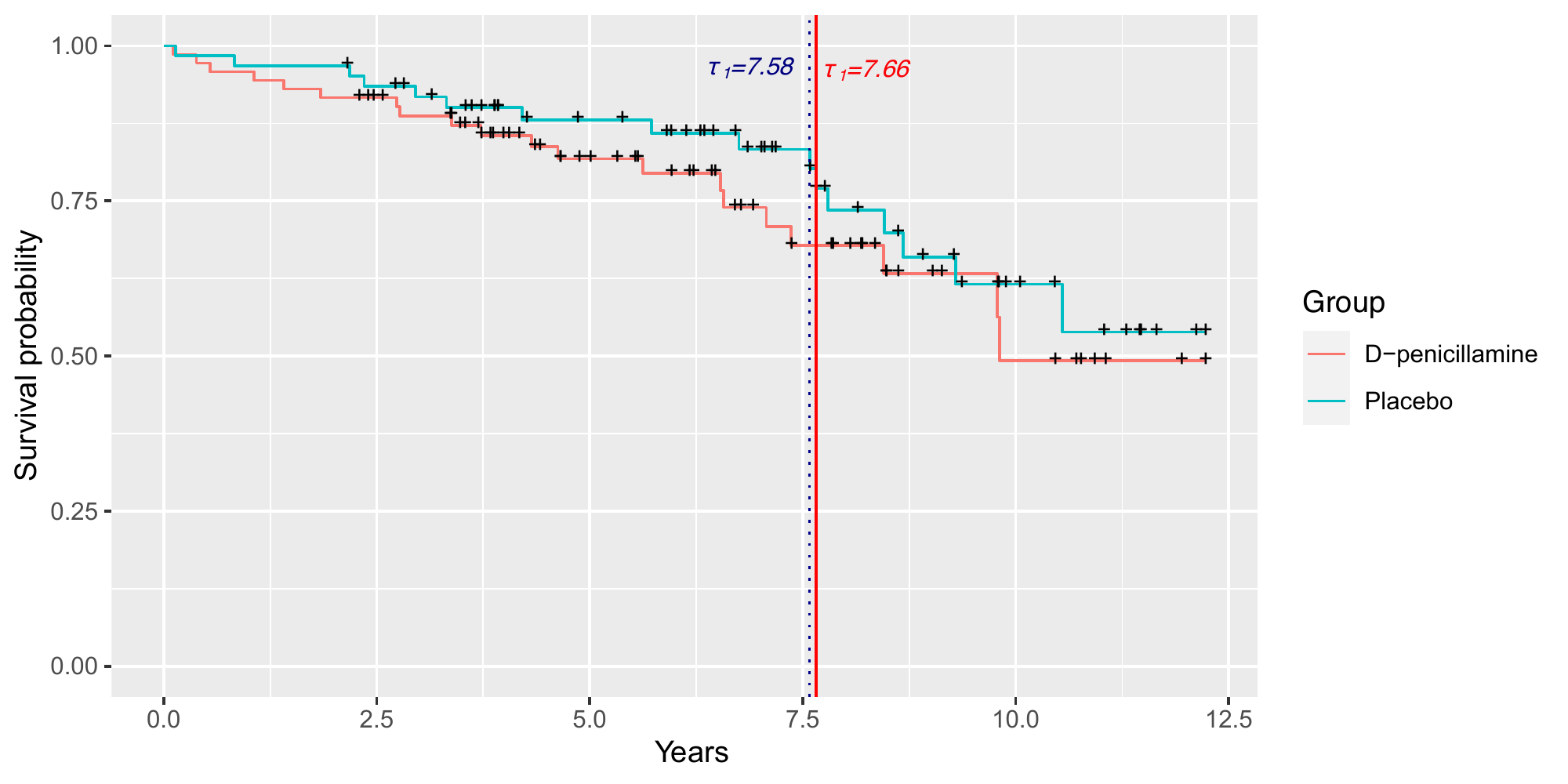}
  \caption{Survival curves of cases in the PBC study}
  \label{PBC}
        \footnotesize{$+$ means censored. The red line and red letters are the change points of the frailty model with one change point. The dark blue dotted line and dark blue letters are the change points of the Cox proportional hazard with one change point.}
  \end{center}
\end{figure}

We included data from a clinical trial on primary biliary cholangitis (PBC); this was a placebo-controlled randomized trial that included 72 patients in the D-penicillamine group and 62 patients in the placebo group\cite{neuberger1985double}. There is stage (S1: no progression, S2: mild progression, S3: moderate progression, and S4: advanced progression) as the cluster. In the D-penicillamine group, the number of S1 is $9$, the number of S2 is $26$, the number of S3 is $29$, and the number of S4 is $8$. In the placebo group, the number of S1 is $3$, the number of S2 is $18$, the number of S3 is $32$, and the number of S4 is $9$. The survival curves are shown in Figure \ref{PBC}. The survival curves crossed near the 10-year time point.
For the Cox proportional hazard model with one change point, $\bbeh=(0.61,-0.54)^T$ and $\tau_1=7.58$. For the frailty model with one change point, $\bbeh=(0.73,-0.19)^T$, $\tau_1=7.66$, and $\bthh=(0.66,1.70)^T$. The data before the change point had a small random effect, but after the change point, the random effect was large and the value of $\beh_2$ changed depending on the change point. The change point did not change markedly, but it shifted back a bit to a point in time by taking into account the random effect.
\subsection{Re-analysis of data of a clinical trial on malignant glioma}
\begin{figure}[H]
  \begin{center}
  \includegraphics[width=15cm]{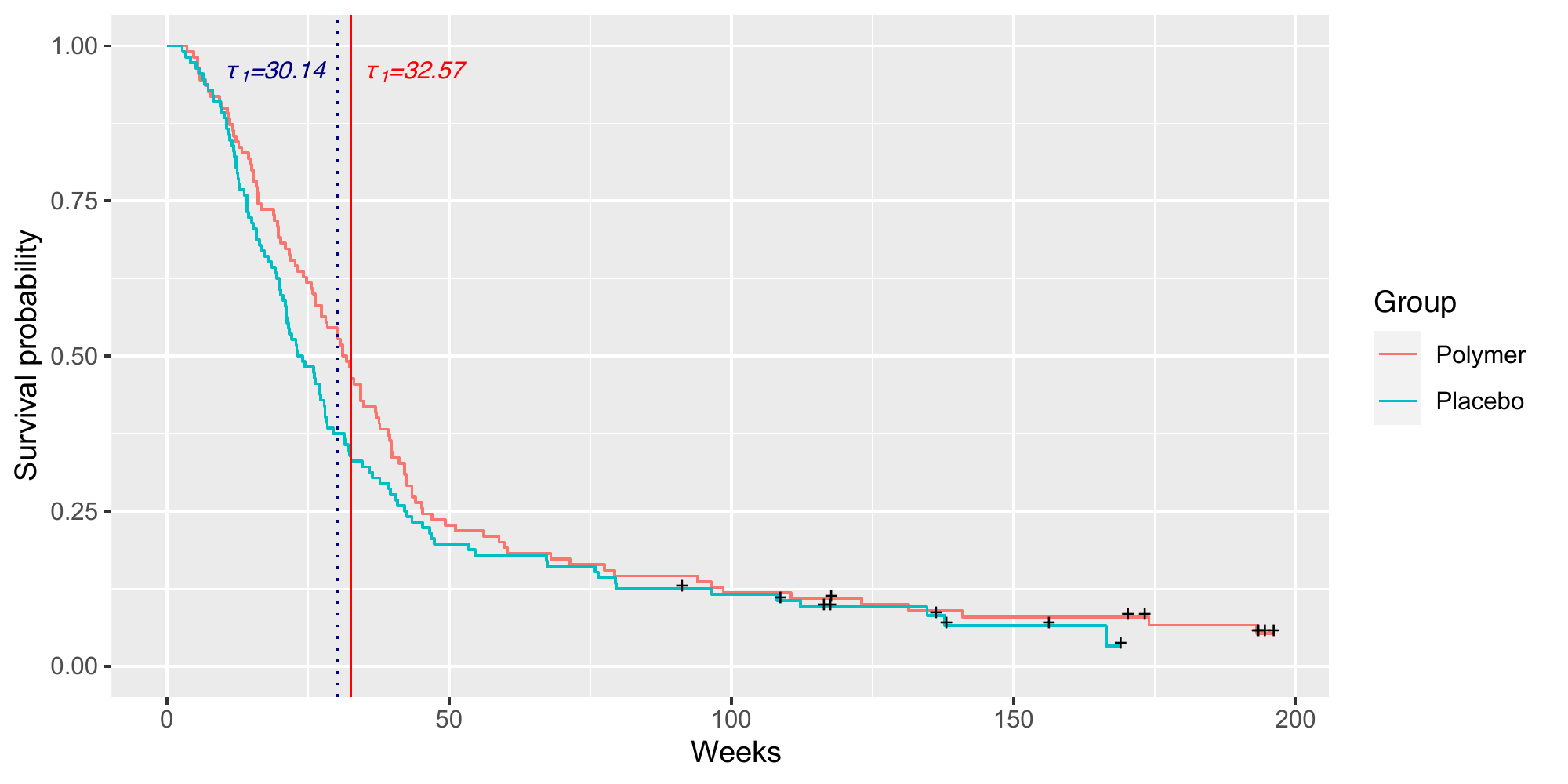}
  \caption{Survival curves of cases in the MG study}
  \label{MG}
        \footnotesize{$+$ means censored. The red line and red letters are the change points of the frailty model with one change point. The dark blue dotted line and dark blue letters are the change points of the Cox proportional hazard with one change point.}
  \end{center}
\end{figure}

We included data from a clinical trial on malignant glioma (MG); the placebo-controlled randomized trial contained 110 patients in the group of patients treated with chemotherapeutic agents incorporated into biodegradable polymers (polymer) and 112 patients in the placebo group\cite{brem1995placebo}. There is tumor histopathology at implementation (path) (P1: glioblastoma, P2: anaplastic astrocytoma, P3: oligodendroglioma, P4: other) as the cluster. For the polymer group, the number of P1 is $76$, the number of P2 is $14$, the number of P3 is $15$, and the number of P4 is $5$. For the placebo group, the number of P1 is $73$, the number of P2 is $16$, the number of P3 is $20$, and the number of P4 is $3$. The survival curves are shown in Figure \ref{MG}.
For the Cox proportional hazard model with one change point, $\bbeh=(-0.44,0.18)^T$ and $\tau_1=30.14$. For the frailty model with one change point, $\bbeh=(-0.45,0.18)^T$, $\tau_1=32.57$, and $\bthh=(0.40,0.08)^T$. Because the random effects are small, the estimators remained almost unchanged.
\section{Discussion}
We propose the novel frailty model with change points to adjust the heterogeneity between clusters. There are clusters of primary diseases, clinical facilities, severity of interest disease, and so on. Our proposed model can be easily analyzed using the coxph function and frailty option. We confirmed that the accuracy of the estimation is increased by considering heterogeneity. Our simulation studies showed that the accuracy of change point estimation deteriorated because of heterogeneity, and the accuracy of estimation was improved using the frailty model. We included the R program code of analyses for the practical example, simulation, and two clinical trials in the Supplementary Materials.
From the simulation study, we confirmed that adding frailty in all estimators for all scenarios resulted in smaller MSEs. Increasing the size of the random effect confirmed that the estimated change point of the Cox proportional hazard model with the change point model has a larger MSE. This suggests that the random effect affects the estimation accuracy of the change point. In the frailty model with the change point, the MSE of the change point estimator became smaller as the random effect increased. The accuracy of data estimation increased as the random effects were adjusted. The bias of the average estimated change points is larger for the frailty model, but the MSE is very large that we considered the difference in bias between the models to be within the margin of error.
In the re-analysis of data from the PBC trial, the random effect size was more significant in the interval where the survival function is crossed. We confirmed that the estimates of the parameters of the regression coefficients change when the random effect is taken into account. In the re-analysis of data from the MG trial, the estimation results did not differ regardless of the presence or absence of frailty because of the small variate effects. Thus, this analysis confirmed that our model does not deviate from the estimation results of the Cox proportional hazard model with a change point when the variate effects are small.
In conclusion, we confirmed that the frailty model with change points has higher accuracy than the model without the random effect. Our proposed model can be easily analyzed using the existing R package. Our proposed model is useful when heterogeneity needs to be taken into account. Additionally, the absence of heterogeneity did not affect the estimation of the regression coefficient parameters.

\bigskip
\noindent
{\bf Author Contributions.}
MK analyzed the example data, simulation datasets, and two clinical trial data and wrote the manuscript. OS proposed the research theme, reviewed and corrected the manuscript.

\bigskip
\noindent
{\bf Acknowledgements.}
MK would like to thank Associate Professor Hisashi Noma for his encouragement and helpful suggestions. The authors would like to thank PhD student Ryoto Ozaki for his helpful comments.
\bibliography{main.bib} 
\bibliographystyle{unsrt} 
\end{document}